\documentclass[a4paper,fleqn]{cas-dc}
\ExplSyntaxOn
\exp_args:NNno \exp_args:Nno \use:n { \cs_gset:Npn \__make_fig_caption:nn #1#2 }
{
	\exp_after:wN \use_ii_i:nn \exp_after:wN
	{ \__make_fig_caption:nn {#1} {#2} }
	{ \dim_set:Nn \l_fig_width_dim \linewidth }
}
\exp_args:NNno \exp_args:Nno \use:n { \cs_gset:Npn \__make_tbl_caption:nn #1#2 }
{
	\exp_after:wN \use_ii_i:nn \exp_after:wN
	{ \__make_tbl_caption:nn {#1} {#2} }
	{ \dim_set:Nn \l_tbl_width_dim \linewidth }
}
\ExplSyntaxOff

\usepackage{lineno,hyperref}
\usepackage[numbers]{natbib}
\usepackage{caption} 
\usepackage{float}  
\usepackage{balance} 
\usepackage{graphicx}  
\usepackage{hyperref}
\usepackage{xcolor}
\modulolinenumbers[5]
\def\tsc#1{\csdef{#1}{\textsc{\lowercase{#1}}\xspace}}
\tsc{WGM}
\tsc{QE}
\tsc{EP}
\tsc{PMS}
\tsc{BEC}
\tsc{DE}

\begin{document}
	\sloppy
	\let\WriteBookmarks\relax
	\def\floatpagepagefraction{1}
	\def\textpagefraction{.001}
	\shorttitle{Neuroaffective generation}
	\shortauthors{Haidong Wang et~al.}
	
	\title [mode = title]{An Audio-Visual Fusion Emotion Generation Model Based on Neuroanatomical Alignment}                      
	\tnotemark[1,2]
	
	\tnotetext[1]{The authors gratefully acknowledge the financial support provided by the Natural Science Foundation of Hunan Province (No. 2024JJ6190), and the Open Project of Xiangjiang Laboratory (No. 23XJ03009). Additional support was provided by the Xiangjiang Laboratory major program subproject (No. 22XJ01001-2).}
	
	\tnotetext[2]{This work also received funding from the "Digital Intelligence +" interdisciplinary research project of Hunan University of Technology and Business (No. 2023SZJ19), as well as from a scientific research project of the Hunan Provincial Department of Education (No. 22B0646).}

	\author[1,2]{Haidong Wang}[type=editor, auid=000, bioid=1, prefix=Sir, role=Researcher, orcid=0000-0002-4614-5817]
	\cormark[1]
	\fnmark[1]
	\ead{whd@hutb.edu.cn}
	\ead[url]{https://github.com/donghaiwang}
	\credit{Conceptualization of this study, Methodology, Software}
	\address[1]{School of Computer Science, Hunan University of Technology and Business, Changsha 410205, 
		China}
	\address[2]{Xiangjiang Laboratory, Changsha 410205, China}
	
	\author[1]{Qia Shan}[style=chinese]
	\fnmark[1]
	\ead{S540534349@163.com}
	\ead[url]{https://github.com/shanqia01}
	\credit{Data curation, Writing - Original draft preparation}
	
	\author[1]{JianHua Zhang}[style=chinese]
	\credit{Data analysis, Writing - Review & Editing}
	
	\author[1]{PengFei Xiao}[style=chinese]
	\credit{Investigation, Formal analysis}
	
	\author[1]{Ao Liu}[style=chinese]
	\credit{Validation, Resources}
	
	\cortext[cor1]{Corresponding author: Haidong Wang}
	\fntext[fn1]{Haidong Wang is the supervising professor and provided guidance throughout the study. Jian Hua Zhang also contributed to conceptualization and project management.}
	\fntext[fn2]{Qia Shan was responsible for the data curation and writing of the original draft. They also coordinated the work between the team members and were involved in reviewing and editing the manuscript.}
	

	\begin{abstract}
		In the field of affective computing, traditional methods for generating emotions predominantly rely on deep learning techniques and large-scale emotion datasets. However, deep learning techniques are often complex and difficult to interpret, and standardizing large-scale emotional datasets are difficult and costly to establish. To tackle these challenges, we introduce a novel framework named Audio-Visual Fusion for Brain-like Emotion Learning(AVF-BEL). In contrast to conventional brain-inspired emotion learning methods, this approach improves the audio-visual emotion fusion and generation model through the integration of modular components, thereby enabling more lightweight and interpretable emotion learning and generation processes. The framework simulates the integration of the visual, auditory, and emotional pathways of the brain, optimizes the fusion of emotional features across visual and auditory modalities, and improves upon the traditional Brain Emotional Learning (BEL) model. The experimental results indicate a significant improvement in the similarity of the audio-visual fusion emotion learning generation model compared to single-modality visual and auditory emotion learning and generation model. Ultimately, this aligns with the fundamental phenomenon of heightened emotion generation facilitated by the integrated impact of visual and auditory stimuli. This contribution not only enhances the interpretability and efficiency of affective intelligence but also provides new insights and pathways for advancing affective computing technology. Our source code can be accessed here: \href{https://github.com/OpenHUTB/emotion}{https://github.com/OpenHUTB/emotion.}
	\end{abstract}

	\maketitle
	
	\section{Introduction}
	
	Emotions are fundamental to human life, serving as one of the complex brain functions crucial for communication, adaptation, and survival~\cite{panksepp1998foundations,zhu2023multimodal}. In human communication, emotions are crucial and represent a key aspect of human intelligence~\cite{massey2002brief,surakka1998modulation}. Sentiment analysis has gained significant attention in artificial intelligence and natural language processing fields~\cite{gandhi2023multimodal}. Affective computing combines research in emotion recognition and sentiment analysis and can utilize either unimodal or multimodal data. This data primarily includes physical information, such as text, audio,  visual and remote photoplethysmography signals (rPPG)~\cite{wang2022systematic,casado2023depression}.
	
	The study of brain and emotion science traces back to 1884 with the introduction of the first explicit theory of emotions, known as the James-Lange theory~\cite{cannon1927james}. According to this theory, emotional experiences stem from responses to physiological changes in the body, where these physiological changes themselves constitute the emotion. During the 1930s, the Papez circuit theory was introduced, proposing that emotional experiences are shaped by the activity of the cingulate cortex and indirect activity from other cortical regions~\cite{papez1937proposed}. MacLean's research suggests that the evolution of the limbic system allowed animals to experience and express emotions, liberating them from stereotyped behaviors dominated by the brainstem~\cite{maclean1952some}. Beau suggests that the anterior superior temporal gyrus houses both visual and auditory areas, utilizing shared neural codes for emotion perception~\cite{sievers2021visual,chua2022predicting}.

	The thalamus, sensory cortex, anterior superior temporal gyrus, amygdala, and orbitofrontal cortex are all involved in emotional learning and expression. These structures collectively comprise an emotional circuit that effectively processes information from sensory organs across different modalities, enabling the learning and expression of emotions. In the context of rapid advancements in artificial intelligence, Morén and Balkenius introduced the Brain Emotional Learning(BEL) model~\cite{moren2001emotional}, which aims to elucidate the neurophysiological mechanisms underlying emotion formation and learning in the mammalian brain by simulating emotional processing between the amygdala and orbitofrontal cortex. Subsequently, researchers integrated the BEL model with artificial neural networks, leading to significant applications in pattern recognition, intelligent control, and facial expression classification~\cite{babaie2008learning,ghanbari2012brain,farooq2020online}. However, research on the mechanisms of emotional learning and generation in the brain remains limited, especially in terms of improving the BEL model to better simulate human emotional learning and generation, which is crucial for advancing affective computing.
	
	In this paper, we present a novel neuroanatomically aligned audio-visual fusion brain emotion learning and generating architecture. The traditional BEL model is an artificial intelligence framework inspired by the emotional processing mechanisms of the brain, primarily based on the interactions between the limbic system of mammals, particularly the amygdala and the orbitofrontal cortex. However, it adopts a simplistic unimodal approach to simulate sensory cortex functions, overlooking essential neurological mechanisms. Our biomimetic emotional pathway is more comprehensive than traditional BEL models, as it integrates the functional of both visual and auditory pathways to achieve advanced biomimicry and interpretability features. 

\begin{figure}[h]
		\centering
		\includegraphics[width=\linewidth]{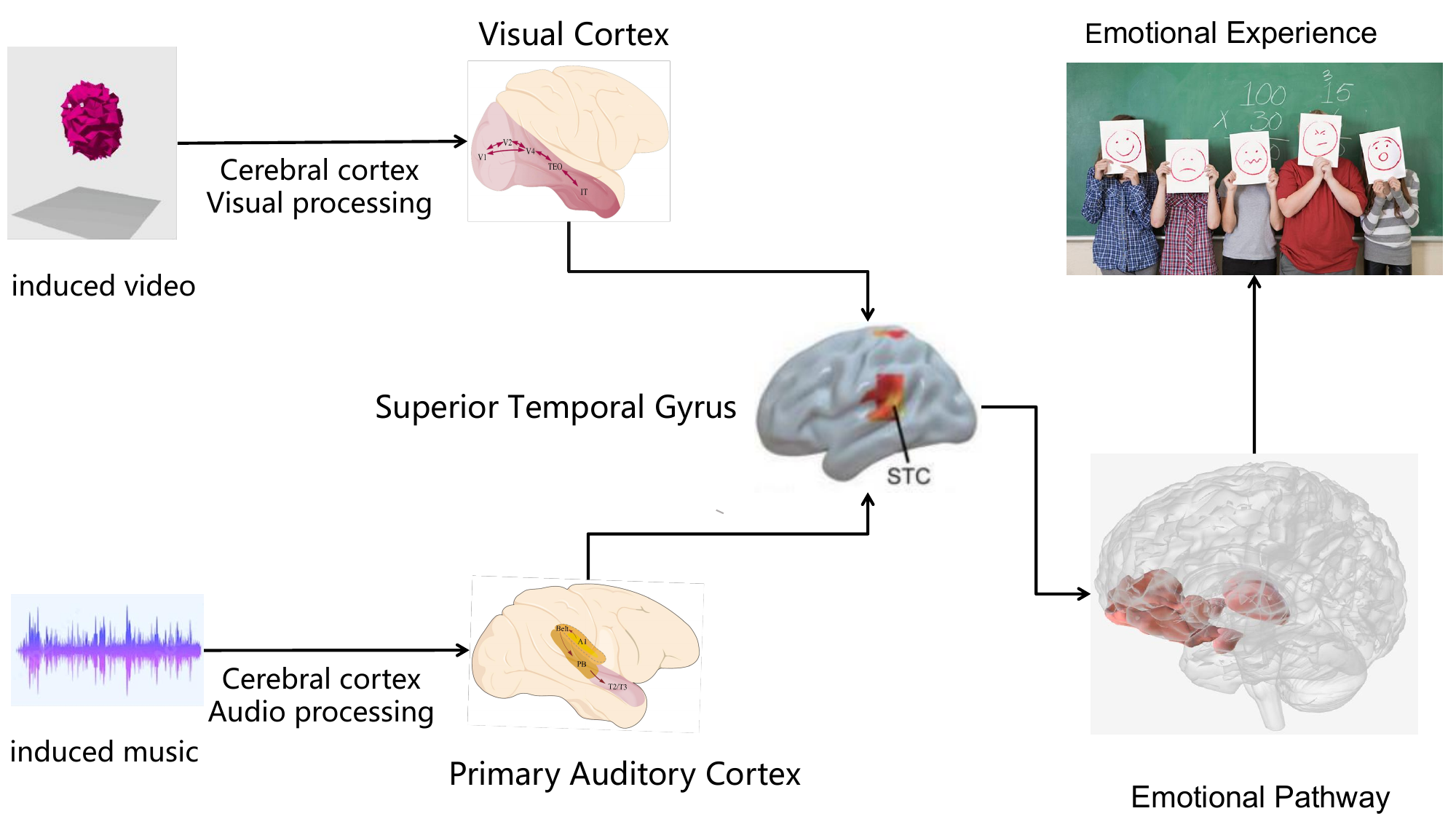}
		\caption{The brain emotional learning and generation pathway of audio visual integration}
		\label{fig:1}
	\end{figure}

	As illustrated in Fig \ref{fig:1}, visual and auditory emotional stimuli are processed into emotional signals through the visual and auditory cortex modules respectively. These signals then undergo a gradual emotional transformation within the emotional pathway of the brain. This pathway mainly includes emotional processing between the visual cortex, primary auditory cortex, anterior superior temporal gyrus, amygdala, and orbitofrontal cortex. Together, they facilitate the emotional function of the brain, translating visual and auditory stimuli into emotional experiences. Our primary objective is to engage in biomimetic modeling of this pathway, with the aim of simulating its functionality while ensuring that the biomimetic modules closely align with the neural structures of the brain. In summary, our main contribution is the introduction of a novel audio-visual fusion brain emotion learning model. This model is designed with neuroanatomical alignment and tailored for affective computing to facilitate lighter and more interpretable artificial intelligence emotion generation.
		
	\section{Related Work}
	\subsection{Brain emotional learning}
	The emotional circuits in the brain have long been a central focus of neurophysiological research, as understanding the mechanisms behind emotions is crucial for motivations related to human-machine interaction and artificial intelligence in the context of hybrid intelligence~\cite{erol2019toward,casado2023depression}. Similarly, simulating the functions of emotional pathways in the human brain is a prominent area of study in the fields of artificial intelligence (AI) and affective computing, where researchers aim to replicate or model the emotional responses and behaviors of the brain. 
	
	In this context, Morén et al. proposed a BEL model, which was inspired by the organizational structure of the limbic system in the mammalian brain. The BEL model emphasizes the interaction between key brain regions involved in emotional processing, such as the amygdala and the orbitofrontal cortex. It simulates the transmission and processing of emotional stimuli through the reflex pathways of the brain, offering insights into how emotional information is processed and reacted to by the neural circuits of the brain~\cite{moren2001emotional,moren2004emotion}. Building upon this foundational work, Lotfi et al. further developed a supervised version of the Brain Emotion Learning model. This enhanced model is capable of learning from target pattern examples, and has been experimentally compared with several alternative approaches, including Multi-Layer Perceptron (MLP) and Adaptive Neuro-Fuzzy Inference Systems (ANFIS). The distinctive feature of the supervised BEL model lies in its ability to achieve rapid training for predictive problems, making it highly effective for real-time applications in emotional learning and decision-making~\cite{lotfi2012supervised}. In a further extension of the BEL framework, Wang et al. proposed a biomimetic memory circuit with both emotion learning and generation capabilities. This circuit is designed to conduct neuromorphic emotion learning and generation, enabling the processing of various input types to simulate emotional responses. Their model advances the state of emotional circuit simulation by incorporating memory elements, which are essential for emulating the ability of the brain to recall and modify emotional states based on past experiences~\cite{wang2021memristive}. Sun et al.                                                                                                                                                                                                                                                                                                                      introduced a biomimetic circuit that simulates a three-dimensional emotional space model. This circuit generates brain-like emotional responses based on multimodal input, including visual, speech, and text information. This model represents a significant step forward in the development of emotion-sensing and emotion-generating systems for bionic robots, providing a potential framework for achieving emotional companionship between humans and machines~\cite{sun2023memristor}. 
	
	While the BEL model has made notable progress in the domains of intelligent control, classification prediction, and the development of emotional circuits, the simulation and utilization of the emotional circuitry of the brain have encountered certain bottlenecks in recent years. With ongoing advancements in neuroscience and neuroengineering, it has become increasingly important to refine and enhance the biomimetic simulation of emotional circuits to improve their functionality, accuracy, and adaptability. The proposed AVF-BEL model represents an advancement over the original BEL framework, not only improving the biomimetic capabilities of emotional circuitry but also offering significant benefits in terms of design efficiency, lightweight operation, and interpretability. This development paves the way for more complex models capable of processing complex emotional stimuli in a manner more akin to the brain, potentially addressing some of the current limitations. It is hoped that, in the future, more effective emotion simulations can be realized in artificial intelligence systems and robotic applications.
	
	\subsection{Emotion generation technology}
    Emotion generation techniques utilize multimodal data, such as images, text, and music, to enable computers to simulate or generate emotional responses. This multidisciplinary approach leverages diverse sources of input to capture the complexity of emotional states and enhance the emotional intelligence of computational systems. 
    
    Zhou et al. introduced an innovative model called the Emotional Chat Machine (ECM), which not only responds appropriately in terms of content relevance and grammatical structure but also effectively manages the emotional aspects of dialogue, ensuring emotional consistency throughout interactions. The ECM model represents a pioneering attempt to address the integration of emotional elements within large-scale dialogue generation systems. Experimental results demonstrate that the model can generate responses that are both contextually appropriate and emotionally congruent, highlighting its potential for enhancing affective communication in automated systems~\cite{zhou2018emotional}. Building on this work, Wang et al. proposed a novel deep generative model designed to synthesize facial videos based on neutral facial images and specific facial expression labels, such as spontaneous smiles. The model is composed of two main components: an image generator and a frame sequence generator. The image generator is realized through a deep neural network that combines Generative Adversarial Networks (GANs) with Variational Autoencoders (VAEs), while the sequence generator is a label-conditioned recursive neural network. This approach allows for the generation of dynamic facial expressions that are temporally consistent. Experimental results validate the models effectiveness, demonstrating the advantages of integrating adversarial elements into the recurrent architecture for facial video synthesis, thus enhancing the realism and expressiveness of the generated videos~\cite{wang2020learning}. In another advancement, Chen et al. proposed a memory-based emotion generation circuit, which is capable of storing emotional memories, retrieving them as needed, and deploying them to generate consistent personality traits. This circuit is designed to address the challenges of low power consumption, area efficiency, and memory processing, making it suitable for integration into resource-constrained systems, such as mobile robots or embedded devices. The ability to store and recall emotional memories is crucial for enabling more natural and adaptive emotional responses in machines, reflecting the dynamic nature of human emotion processing~\cite{chen2023memristive}. Further contributing to the field, Zhang et al. introduced a brain-inspired emotion generation system based on the Pleasure-Arousal-Dominance (PAD) model, a three-dimensional emotional space that simulates the neural circuits involved in emotion generation within the limbic system. This model draws inspiration from the biological structures of the brain and provides a highly realistic and biologically grounded approach to emotion simulation. The PAD model offers valuable insights into the neural underpinnings of emotion and sets the stage for the development of more accurate and nuanced emotional simulations in artificial systems~\cite{zhang2024memristive}. 
    
    Advancements in emotion generation technologies have made significant progress in areas such as text-based dialogue systems, emotion generation circuits, and facial expression synthesis. These breakthroughs rely heavily on sophisticated deep learning techniques, extensive emotional datasets, and optimized hardware components. However, despite these advancements, widespread deployment and application of emotion generation technologies remain challenging due to the substantial investments required in both software and hardware infrastructure. As such, there are significant barriers to achieving scalable and cost-effective solutions. In contrast to many of these resource-intensive approaches, our AVF-BEL model emphasizes a lightweight design and high interpretability, while also achieving excellent performance in neuroanatomical alignment and result similarity. These features are particularly valuable for large-scale applications in areas such as robot emotion generation and artificial intelligence emotion synthesis. By optimizing for efficiency and ease of interpretation, the AVF-BEL model offers a promising avenue for overcoming the challenges associated with resource-intensive emotion generation systems, potentially enabling more widespread adoption and practical implementation in real-world applications.
	
	\section{Method}
	\subsection{Biological mechanism}
	\label{subsec:3.1}
	
	In developing the AVF-BEL framework, our primary inspiration stemmed from four neuroscientific brain structures:visual pathway, primary auditory cortex, anterior superior temporal gyrus, emotional pathway. Specifically, these four neural structures within neuroscience form an integrated pathway for emotional learning and generation. This pathway combines visual and auditory stimuli received by sensory cortices to shape emotional experiences. 
	
	As illustrated in Fig \ref{fig:1}, the referenced emotional pathway represents the short emotional circuit in the brain. It primarily involves emotional interactions between the amygdala and orbitofrontal cortex, both of which are key to emotion regulation and learning, with characteristics of simplification and biomimetic features~\cite{moren2001emotional,rolls2019cingulate,birnie2022principles}.
	
	\textbf{\textsf{Visual Pathway:}} The discovery and analysis of cortical visual areas represent a significant achievement in visual neuroscience~\cite{grill2004human,gu2019modality}. The visual cortex, located in the occipital lobe at the posterior region of the cerebral cortex, serves as the principal region of the brain for receiving, integrating, and processing visual information from the retina. The primary visual cortex (V1) is the initial cortical area engaged in this processing, where basic visual features such as edges, orientation, and spatial frequency are analyzed.
	
	From V1, visual information progresses along distinct pathways, each specializing in different aspects of visual processing. The ventral stream (V1 → V2 → V4 → IT) primarily processes object recognition and identification, analyzing complex shapes, colors, and textures to support higher-level visual cognition in regions such as the inferior temporal cortex (IT) ~\cite{huff2018neuroanatomy,siu2018development,kubilius2019brain}. The dorsal stream, involving regions such as V1 → V2 → V5/MT (middle temporal) → MST (medial superior temporal), is more specialized for motion processing, spatial perception, and visually guided actions. Areas like V5/MT and MST contain motion-sensitive neurons with larger receptive fields, enabling spatial integration and the detection of motion direction and speed~\cite{furlan2016global,koelsch2018auditory,campana2006visual}. Each of these pathways within the visual cortex contributes to distinct aspects of visual perception, advancing from initial image processing in V1 to complex visual interpretations necessary for interaction with the environment.
	
	\textbf{\textsf{Primary Auditory Cortex:}} The primary auditory cortex constitutes the core of human auditory capabilities. Understanding the organization of the primary auditory cortex forms the neural foundation for comprehending auditory behavior~\cite{pandya1995anatomy,pekkola2005primary}. It is situated within the lateral sulcus of the brain, specifically within Heschl's gyrus, also referred to as the transverse temporal gyrus. The representational capacity is predominantly dependent on the cortical state, suggesting that the cortical state should be regarded as a prominent variable in all studies of sensory processing. It is essential for the initial processing of auditory information, particularly in distinguishing sound features like frequency, intensity, and duration~\cite{da2011human,pachitariu2015state,park2020circuit}.
	
	\textbf{\textsf{Anterior Superior Temporal Gyrus:}} The superior temporal sulcus (STS) is the primary region for audio-visual integration. It is vital for perceiving biological motion and is also considered essential for speech and facial processing~\cite{hein2008superior}. Within the superior temporal gyrus (STG), there exists a dual dissociation of auditory components, distinguishing between clear and noisy elements, in response to both auditory and visual speech. This structure is central to the multisensory integration of both auditory and visual speech, even in noisy environments~\cite{ozker2017double}. The multimodal integration of visual and auditory perception in the anterior superior temporal gyrus enhances the speed and accuracy of emotional information processing. Located in the lateral sulcus of the temporal lobe, between the lateral fissure and superior temporal sulcus, this area is crucial for the integrated perception of visual and auditory emotions in the brain~\cite{gao2020fmri,sievers2021visual}.
	
	\textbf{\textsf{Emotional Pathway:}} Evidence from anatomy, neurophysiology, functional neuroimaging, and neuropsychology delineates the anterior margin and associated structures. This encompasses the orbitofrontal cortex and amygdala, which are implicated in emotion, reward evaluation, and reward-related decision-making (excluding memory). Value representations are channeled to the anterior cingulate cortex for learning action outcomes. Various limbic structures, encompassing the amygdala and orbitofrontal emotional system, exhibit distinct connectivity and functionality~\cite{rolls2015limbic,kragel2016decoding}. 
	
	\subsection{Architecture design}
	\label{subsec:subsection0}
	
\begin{figure}[h]
		\centering
		\includegraphics[width=\linewidth]{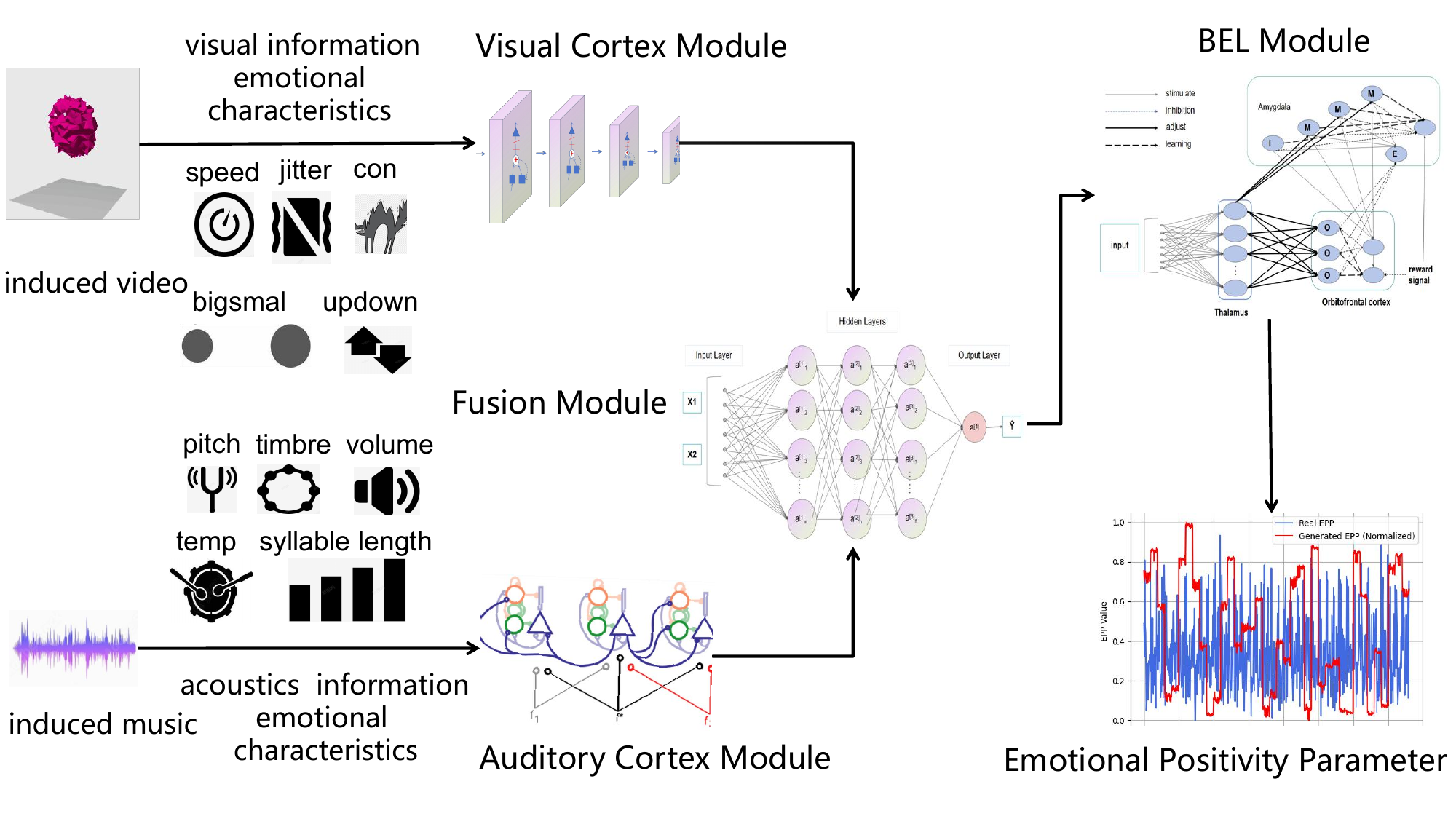}
		\captionsetup{justification=raggedright, skip=10pt} 
		\caption{AVF-BEL model framework:The visual cortex module is based on the modified CORnet-Z model, while the auditory cortex module utilizes a simplified primary auditory cortex model, Col1\_fs. The fusion module uses an audiovisual integration MLP, while the BEL module refines sensory cortex input, building on the traditional BEL model.}
		\label{fig:2}
	\end{figure}

	As demonstrated in Fig \ref{fig:2}, we have designed and implemented a comprehensive brain-based emotional learning and generation framework that facilitates audio-visual integration. This architecture is structured to mimic the processing of emotional stimuli in the brain, with a focus on visual and auditory modalities. Initially, visual stimuli are processed sequentially within the visual cortex module, where various visual features related to emotional content, such as color, motion, and spatial orientation, are analyzed and encoded. Simultaneously, auditory stimuli are processed within the auditory cortex module, where acoustic features, including pitch, rhythm, and tone, are extracted and mapped to emotional relevance. The next critical step involves the fusion module, where information from both the visual and auditory modalities is integrated. This module facilitates the synthesis of multi-sensory inputs, creating a cohesive representation of the emotional content derived from both visual and auditory cues. The integrated emotional features are then passed on to the BEL module, which performs further processing and learning. This module is designed to simulate the neurobiological mechanisms responsible for transforming visual and auditory stimuli into emotional parameters, specifically focusing on the generation of emotional positivity. 
	 
	One of the key innovations of our model is its alignment with neuroanatomical structures involved in emotional processing. The architecture is carefully designed to reflect the anatomical and functional interconnections between the visual cortex, auditory cortex, and the emotional learning structures of the brain. By incorporating these four core modules—visual cortex, auditory cortex, fusion, and BEL—our model not only replicates the process of emotional perception and generation but also achieves a high level of neuroanatomical alignment. This design ensures that the operation of the model is grounded in a biologically plausible framework, enhancing its validity and interpretability.
       
   \begin{figure*}[htbp]
   	\centering
   	\includegraphics[width=\linewidth]{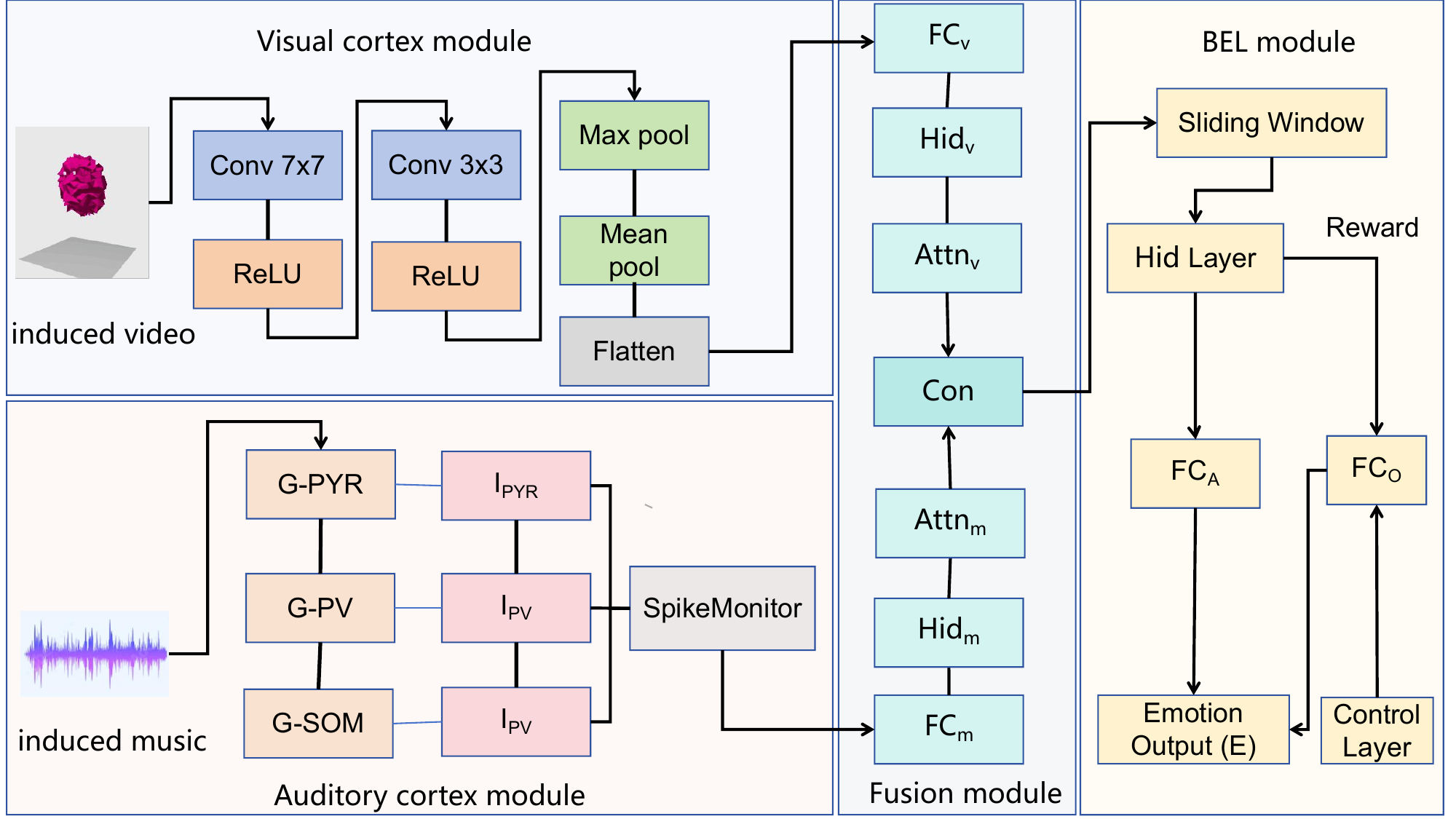}
   	\captionsetup{justification=raggedright, skip=10pt} 
   	\caption{Neural network architecture diagram for emotion generation with integrated visual and auditory processing for dynamic emotional response.}
   	\label{fig:3}
   \end{figure*}
   
    As demonstrated in Fig \ref{fig:3}, the audiovisual emotion generation system operates through multi-module collaboration, enabling the extraction, integration, and analysis of visual and auditory features. The visual cortex module processes emotional visual features based on an enhanced architecture. The input layer receives five emotion-related visual features, with the first convolutional layer using a large 7$\times$7 kernel to rapidly reduce input dimensions, followed by successive layers with 3$\times$3 kernels to refine features incrementally. Common activation functions include Sigmoid, GELU and ReLU. The unidirectional activation property of ReLU resembles the ``on/off'' behavior of biological neurons, where activation occurs only when the input signal exceeds a certain threshold. This simplifies the computational process, accelerates gradient propagation, and avoids delays caused by complex computations. Additionally, the sparsity of ReLU helps reduce overfitting and improves computational efficiency, while effectively mitigating the vanishing gradient problem, ensuring stable gradient updates within the neural network~\cite{he2024text}. A max-pooling layer reduces spatial dimensions, and an adaptive average pooling layer reduces each feature map to a 1x1 size, extracting a single feature value per channel. The flattened layer transforms the features into a one-dimensional vector, which is then mapped to the emotional output space by a linear classifier. The auditory cortex module simulates auditory processing using a simplified neural model, defining dynamic equations for excitatory and inhibitory neurons-G-PYR (Generate pyramid), G-PV (Generate parvalbumin), and G-SOM (Generate somatostatin)—while mapping five acoustic feature input currents onto each neuron, which in turn generate auditory emotional features. The fusion module employs cross-modal fusion to process and integrate audio and visual features. It first uses fully connected layers ($\mathit{FC_{\text{m}}}$ and $\mathit{FC_{\text{v}}}$) to map the audio and visual inputs into their respective hidden feature spaces ($\mathit{Hid_{\text{m}}}$ and $\mathit{Hid_{\text{v}}}$). Attention mechanisms ($\mathit{Attn_{\text{m}}}$ and $\mathit{Attn_{\text{v}}}$) then calculate attention weights, dynamically adjusting the importance of each modality. These weighted features are concatenated and passed through a final fully connected layer before being fed into the subsequent BEL module. The BEL module utilizes a recurrent neural network architecture to process and integrate audiovisual emotional inputs received from the fusion module. First, temporal features are extracted using a sliding window approach to capture time dependencies. The hidden layer stores the learned contextual information, which is then passed through fully connected layers to generate excitatory signals ($\mathit{FC_{\text{A}}}$) and inhibitory signals ($\mathit{FC_{\text{O}}}$). The control layer mimics the prefrontal cortex, fine-tuning these inhibitory signals to ensure appropriate emotional responses. A reinforcement signal (Reward) provides feedback to optimize the learning of the system and strengthen desirable emotional behaviors. The following sections provide a detailed overview of the specific roles and functions of each of these four modules within the overall emotional learning and generation process.
    
	\textbf{\textsf{Visual Cortex Module:}} Kubilius et al. developed the CORnet series of models, a shallow artificial neural network (ANN) featuring four anatomical mapping regions and recurrent connections, guided by Brain-Score~\cite{kubilius2018cornet}. This represents a large-scale amalgamation of novel neural and behavioral benchmarks, designed to assess the functional fidelity of ventral visual stream models in primates. Although notably shallower than most models, the CORnet series stands as a top-performing model on Brain-Score, surpassing comparable compact models on ImageNet~\cite{wieser2020understanding,kubilius2019brain}. The visual cortex module is built on an enhanced CORnet-Z model, which is central to extracting and processing visual emotion features for emotion generation tasks~\cite{kubilius2018cornet}. 
	
	Its architecture harnesses the powerful feature extraction capabilities of convolutional neural networks (CNNs), simulating the processing of visual information by the brain by progressively extracting emotion-related visual features across layers. The model consists of convolutional layers, ReLU activation functions, and max-pooling layers. The convolutional layers extract multi-level emotional feature maps from input images, while the ReLU activation introduces non-linearity, enabling the model to capture complex emotional patterns. The max-pooling layers effectively reduce the spatial dimensions of the feature maps, enhancing computational efficiency while preserving essential emotional information. The first convolutional block utilizes a large 7$\times$7 kernel to quickly reduce the input dimensions, followed by subsequent blocks with smaller 3$\times$3 kernels that progressively refine more detailed emotional features. After convolution and pooling, the model uses an adaptive average pooling layer to further reduce the feature maps to a 1$\times$1 size, consolidating each channel into a single feature value. The Flatten layer then transforms these features into a one-dimensional vector, which can be directly accessed through an Identity layer, facilitating subsequent emotional analysis and generation tasks. The weights of the model are initialized using Xavier uniform initialization to ensure stable training, with biases initialized to zero to prevent imbalance in signal propagation during the training process. If BatchNorm layers are included, their parameters are also carefully initialized. The combination of effective dimensionality reduction strategies within the visual processing module enables the model to efficiently extract emotional features from visual data, thereby improving its performance in handling complex emotional imagery. 
	
	The primary function of this module is to input five visual features (speed, jitter, consonance, bigsmall, updown) extracted from original emotional animation videos into the modified CORnet-Z model. The model includes four computational areas that are conceptually analogous to the dorsal visual regions V1, V2, V4, and IT, as well as a linear classifier decoder that maps the neural populations of the final visual area to behavioral choices. In this module, the tensor $\mathit{{i_p}_{\text{n,c,h,w}}}$ is a four-dimensional tensor with dimensions (\textit{n, c, h, w}), where \textit{n} is the batch size, \textit{c} is the number of channels corresponding to the number of features, \textit{h} is the height of $\mathit{{i_p}_{\text{n,c,h,w}}}$, and \textit{w} is the width of $\mathit{{i_p}_{\text{n,c,h,w}}}$. Convolution operations $\mathit{{Conv}_{\text{c,k}}}$ are performed on the input feature tensor $\mathit{{i_p}_{\text{n,c,h,w}}}$ using convolution kernel \textit{k} for each channel, adding a bias term \textit{b}, and then applying the ReLU activation function:
	
	\begin{equation}
		{X_{\text{n,c,h,w}}} = \text{ReLu}( {\textstyle \sum_{k=1}^{c}}({{i_p}_{\text{n,c,h,w}}} \cdot {\text{Conv}_{\text{c,k}}} + {b_{\text{c}}})).
	\end{equation}

	The $\mathit{{Conv}_{\text{c,k}}}$ refers to the convolutional kernels that map $\mathit{c}$ input channels to $\mathit{k}$ output channels. The convolved $\mathit{{i_p}_{\text{n,c,h,w}}}$ is then subjected to max pooling, where the maximum value from each region is selected as the pooling result:
	
	\begin{equation}
		{\hat{X}_{\text{n,c,h,w}}} = \text{maxp}({X_{\text{n,c,h,w}}}).
	\end{equation}
	The flattened and combined tensor $\mathit{X_{\text{n,c,h,w}}}$ is transformed into a one-dimensional vector $\mathit{\hat{X}_{\text{n,c,h,w}}}$. A linear transformation is then applied to the flattened feature vector $\mathit{\hat{X}}$, resulting in the final feature vector $\mathit{X_{\text{a}}}$:
	
	\begin{equation}
		{X_{\text{a}}} = {W \cdot {{\hat{X}}} + b}.
	\end{equation}
	These vectors $\mathit{X_{\text{a}}}$ will be used as input signals for subsequent fusion modules.
	These vectors, $\mathit{X_{\text{a}}}$, will serve as input signals for subsequent fusion modules, where $\mathit{W}$ and $\mathit{b}$ represent the weight matrix and bias term, respectively.
	
	\textbf{\textsf{Auditory Cortex Module:}} The auditory cortex of mammals is comprised of diverse inhibitory and excitatory neuron types, forming intricate microcircuits for the processing and transmission of sensory information. Various subtypes of inhibitory neurons serve distinct roles in auditory processing~\cite{sadagopan2023quantitative}. The auditory cortex module utilizes a simplified primary auditory cortex model, col1\_fs, aimed at simulating the behavior of different types of neurons in the auditory cortex~\cite{chang2021development}. Initially, three populations of neurons are defined (Excitatory pyramidal neurons (Pyr), inhibitory neurons such as parvalbumin expressing neurons (PV) and somatostatin expressing neurons (SOM)), with each population described by differential equations that represent their neural dynamics. The code uses the `NeuronGroup' function to create 400 pyramidal neurons G-PYR, 200 PV neurons G-PV, and 200 SOM neurons G-SOM, establishing a threshold (membrane potential greater than 0.5 volts) and reset conditions (membrane potential returns to zero) for each population. The input currents are set at 0.6 volts for both PYR and PV neurons, and 0.65 volts for SOM neurons, simulating different activation states under realistic physiological conditions. To monitor spiking activity, `SpikeMonitor' is utilized to track the activity of each neuron population. Finally, the simulation is executed for a duration of 1000 milliseconds using `run(1000ms)'. This module provides a foundational framework for studying auditory information processing, enabling the analysis of dynamic behaviors in auditory signal processing through spiking activity monitoring of various neuron populations, thus revealing interactions and functional characteristics among neurons. The primary function of this module is to input five acoustic features (pitch, tonnetz, volume, tempo, duration) extracted from original emotional audio into the modified col1\_fs model~\cite{park2020circuit}. The model consists of an excitatory neural population and two inhibitory neural subpopulations, defined by different types of neuron models: excitatory neurons (G-PYR), inhibitory neurons (G-PV), and SOM inhibitory neurons (G-SOM). Each neuron receives different input currents to simulate neuronal activity in the primary auditory cortex. The dynamic model of G-PYR can be described as:
	
	\begin{equation}
		\frac{dv_{\text{PYR}}}{dt} = \frac{I_{\text{PYR}} - v_{\text{PYR}}}{\tau_{\text{PYR}}},
	\end{equation}
	where $\mathit{v_{\text{PYR}}}$ represents the membrane potential of excitatory neurons (G-PYR). $\mathit{I_{\text{PYR}}}$ is the input current, and $\mathit{\tau_{\text{PYR}}}$ is the time constant.
	The dynamic model of G-PV can be described as:
	
	\begin{equation}
		\frac{dv_{\text{PV}}}{dt} = \frac{I_{\text{PV}} - v_{\text{PV}}}{\tau_{\text{PV}}},
	\end{equation}
	where $\mathit{v_{\text{PV}}}$ represents the membrane potential of inhibitory neurons (G-PV). $\mathit{I_{\text{PV}}}$ is the input current, and $\mathit{\tau_{\text{PV}}}$ is the time constant.
	The dynamic model of G-SOM can be described as:
	
	\begin{equation}
		\frac{dv_{\text{SOM}}}{dt} = \frac{I_{\text{SOM}} - v_{\text{SOM}}}{\tau_{\text{SOM}}},
	\end{equation}
	where $\mathit{v_{\text{SOM}}}$ represents the membrane potential of inhibitory neurons (G-SOM). $\mathit{I_{\text{SOM}}}$ is the input current, and $\mathit{\tau_{\text{SOM}}}$ is the time constant.
	Mapping five emotional auditory features (pitch, tonnetz, volume, tempo, duration) to the input currents of three neurons.
	\begin{equation}
		{I_{\text{PYR}}} = {f_{\text{PYR}}(x_{\text{pi}}, x_{\text{to}}, x_{\text{rm}}, x_{\text{te}}, x_{\text{du}})},
	\end{equation}
	\begin{equation}
		{I_{\text{PV}}} = {f_{\text{PV}}(x_{\text{pi}}, x_{\text{to}}, x_{\text{rm}}, x_{\text{te}}, x_{\text{du}})},
	\end{equation}
	\begin{equation}
		{I_{\text{SOM}}} = {f_{\text{SOM}}(x_{\text{pi}}, x_{\text{to}}, x_{\text{rm}}, x_{\text{te}}, x_{\text{du}})}.
	\end{equation}
	Based on the dynamical model of the neurons described above, it is possible to simulate changes in the membrane potential of the neurons and record their activity states. The functionality of the auditory cortex can be characterized and analyzed by monitoring neuronal activity for feature extraction.
	
	\begin{figure}[h]
		\centering
		\includegraphics[width=\linewidth]{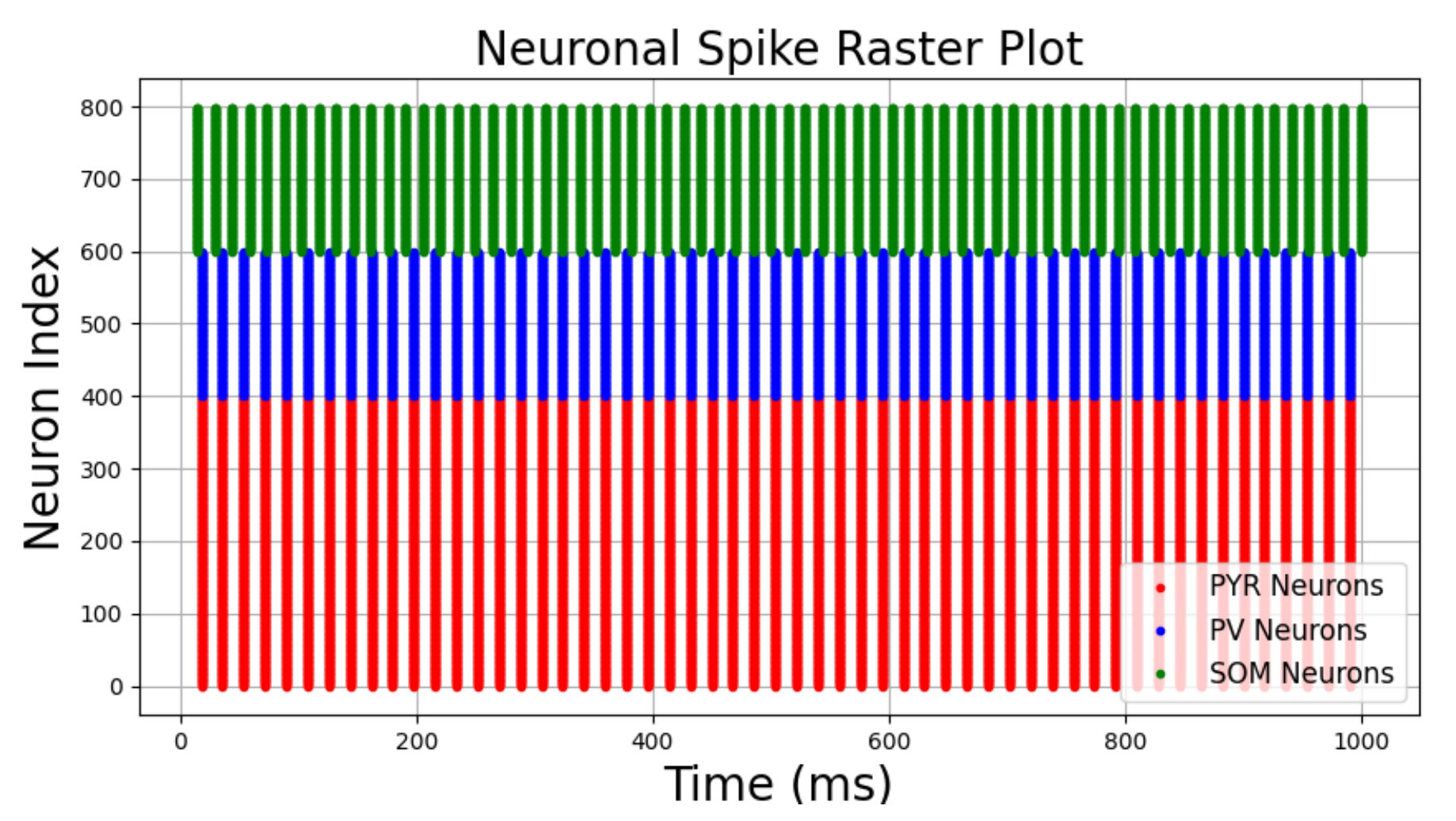}
		\caption{The M-BEL model encompasses three distinct types of neuronal activity, each contributing uniquely to the overall emotional learning and generation process.}
		\label{fig:4}
	\end{figure}
	As illustrated in Fig. \ref{fig:4}, there are 400 PYR neurons, 200 PV neurons, and 200 SOM neurons. spike activity for each type of neuron is monitored over a 1000-millisecond simulation period.The transformation process unfolds as follows:
	\begin{equation}
		{X_{\text{m}}} = {N_{\text{ea}}}({I_{\text{PYR}}}, {I_{\text{PV}}}, {I_{\text{SOM}}}),
	\end{equation}
	where $\mathit{N_{\text{ea}}}$ refers to the simulation and monitoring process of neuronal activity. $\mathit{I_{\text{PYR}}}$, $\mathit{I_{\text{PV}}}$, and $\mathit{I_{\text{SOM}}}$ are functions that map emotional auditory features to the input current of neurons. This equation illustrates the transformation process from emotional auditory features to neuronal activity, demonstrating the complex processing and response of the auditory cortex to musical characteristics.
	
	\textbf{\textsf{Fusion Module:}} The fusion module is designed based on a MLP architecture, specifically tailored to process emotional data from visual and auditory modalities. This module integrates and models features from different modalities through a unified multimodal processing framework. The model takes multimodal data (visual and auditory features) as input, with each modality processed separately through dedicated visual and auditory modules for initial feature extraction. In the architecture design, the input data of each modality first passes through its respective independent MLP branch, each consisting of three fully connected layers combined with ReLU activation functions and batch normalization, effectively capturing intra-modal feature representations. These modality-specific features are then integrated through a cross-modal interaction layer to generate a joint representation. The fused joint feature vector is subsequently passed through fully connected layers to complete the final emotion generation task. The model employs Mean Squared Error (MSE) as the loss function to optimize the prediction of continuous emotional data. The Adam optimizer is utilized with an appropriately set learning rate to balance the trade-off between convergence speed and training stability. To enhance training efficiency and model generalization, the model supports batch loading and random shuffling of multimodal data. The training process is set for 100 epochs, during which the model sequentially extracts visual and auditory features from the batch data in each epoch. These features are forwarded to generate the joint feature representation, with model parameters updated through backpropagation. Training losses can be recorded at fixed intervals (every 10 epochs) for subsequent inference and performance evaluation. With the aforementioned multimodal fusion module design, the model effectively integrates emotional information from both visual and auditory modalities, significantly enhancing the performance and reliability of multimodal emotion analysis, providing robust support for emotion data modeling and analysis.
	
	Overall, this module provides a clear and comprehensive foundation for the joint learning of audio and visual emotional data, making it well-suited for further multi-modal research and applications. This fusion module aims to emulate the function of a singular neural code shared by auditory and visual information within the anterior superior temporal gyrus. The multi-layer structure and non-linear activation functions of the MLP enable it to learn and represent complex non-linear relationships. This capability is particularly important for integrating emotional features, which often exhibit non-linear and intricate characteristics. Additionally, its simple structure allows for seamless integration with the BEL module implemented by recurrent neural network (RNNs), forming a hybrid model that further enhances the performance of emotional learning and generation models. The MLP can effectively integrate emotional features from audio and animation, enabling the simulation and recognition of overall emotional stimuli~\cite{qian2021multi,passos2023multimodal}. In this module, a multimodal Multilayer Perceptron model was implemented, integrating the visual and auditory feature vectors obtained from the preceding two modules. The input feature vectors $\mathit{X_{\text{a}}}$ and $\mathit{X_{\text{m}}}$  are processed by the fully connected layer  and the ReLU activation function to obtain the hidden layer feature vectors $\mathit{H_{\text{m}}}$ and $\mathit{H_{\text{v}}}$:
	
	\begin{equation}
		{H_{\text{m}}} = \text{ReLU}({W_\text{music}({X_{\text{m}}})} + b_{\text{music}}),
	\end{equation}
   where ${W_{\text{music}}}$ is the weight matrix for the audio features. ${b_{\text{music}}}$ is the bias term for the audio features.
	\begin{equation}
		{H_{\text{v}}} = \text{ReLU}({W_{\text{video}}({X_{\text{a}}})}+ b_{\text{video}}),
	\end{equation}
   where ${W_{\text{video}}}$ is the weight matrix for the video features. ${b_{\text{video}}}$ is the bias term for the video features. 
   \begin{equation}
   	{Attn_{\text{m}}} = \text{ReLU}({W_\text{attn-music}({H_{\text{m}}})} + b_{\text{attn-music}}),
   \end{equation}
   where ${W_{\text{attn-music}}}$ is the weight matrix for the audio features. ${b_{\text{attn-music}}}$ is the bias term for the audio features.
   \begin{equation}
   	{Attn_{\text{v}}} = \text{ReLU}({W_\text{attn-video}({H_{\text{v}}})}+ b_{\text{attn-video}}),
   \end{equation}
   where ${W_{\text{attn-video}}}$ is the weight matrix for the video features. ${b_{\text{attn-video}}}$ is the bias term for the video features. 
   The features are then weighted according to the attention weights and fused (concatenated):
   \begin{equation}
   	{X_{\text{m1}}} = {Attn_{\text{m}} \cdot H_{\text{m}} },
   \end{equation} 
  \begin{equation}
	{X_{\text{a1}}} = {Attn_{\text{v}} \cdot H_{\text{v}} }.
  \end{equation}
   Combine the hidden layer feature vectors $\mathit{X_{\text{m1}}}$ and $\mathit{X_{\text{a1}}}$ of the audio and visual into a comprehensive feature vector $\mathit{X_{\text{am}}}$:
	\begin{equation}
		{X_{\text{am}}} = \text{Concat}({X_{\text{a1}}},{X_{\text{m1}}}),
	\end{equation}

	where $\mathit{X_{\text{am}}}$ will be used as input for the next BEL module. This enables seamless integration with the subsequent BEL module implemented using RNNs, thereby further enhancing the performance of the model.
	
	\textbf{\textsf{BEL Module:}} The network design of this module is primarily based on the brain emotion learning model, simulating the emotional interactions between the amygdala and the orbitofrontal cortex. Using the integrated feature vector sequence output from the audio-visual emotion fusion module as input, this module constructs a recurrent neural network based on the BEL model~\cite{moren2001emotional,moren2004emotion,lotfi2013brain,lotfi2018competitive}. The model utilizes a sliding window approach to construct time series data, enabling it to capture dynamic emotional variations. Randomly generated signals are used to simulate thalamic output, facilitating the transmission of dynamic neural signals. The amygdala integrates inputs from the sensory cortex and thalamus, producing excitatory and inhibitory signals that reflect the complexity of emotional decision-making. In terms of the learning system, the orbitofrontal cortex simulates inhibitory functions by integrating inputs from various regions, achieving emotional regulation through behavioral adjustments. During the training process, RNN parameters are defined, and weights are progressively updated through an error feedback mechanism to enhance classification accuracy. Finally, the effectiveness of model evaluation is ensured by calculating accuracy on the test set. 
	
	This module successfully simulates the emotional pathways of the brain through a multi-layered neural network, encompassing critical processes such as visual and auditory information processing, emotional responses, and learning adaptations. The amygdala embodies an excitatory learning system responsible for perceiving and learning inputs. The learning input constitutes part of a negative feedback loop, halting the learning signal once the output attains the reinforcement signal level. Additionally, it receives inhibitory input from the orbitofrontal system, potentially dampening inappropriate emotional responses. The activity of the perception node $\mathit{A}$, corresponding to each stimulus $\mathit{X_{\text{am}}}$, is calculated as follows: $\mathit{A} = \mathit{X_{\text{am}}} \times \mathit{V}$. \textit{V} represents the adaptable connection weight. In the amygdala, reinforcement learning rules are applied to adjust the connection weights \textit{V} of perceptual nodes. This allows the model to adapt its output based on rewards, guiding it towards the desired level of reinforcement. This learning rule can be described by the following formula:
	
	\begin{equation}
		\bigtriangleup V = \alpha \left [ {X_{\text{am}}} \cdot max\left ( 0, {R_{\text{e}}} -  {\textstyle \sum_{0}^{j}A  }  \right )  \right ], 
	\end{equation}
	where $\mathit{\bigtriangleup V} $ represents the adjustment amount of the connection weight \textit{V}. $\mathit{\alpha}$ is the learning rate. $\mathit{X_{\text{am}}}$ represents the comprehensive feature vector from the fusion module. $\mathit{R_{\text{e}}}$ is a reinforcement signal (Reward). $\mathit{\textstyle \sum_{0}^{j}A}$ represents the total output of all perception nodes. This formula indicates that for each perception node, the adjustment amount $\mathit{\bigtriangleup V}$ of its connection weight \textit{V} depends on the discrepancy between the current perceived input $\mathit{X_{\text{am}}}$, the expected reinforcement signal $\mathit{R_{\text{e}}}$, and the current model output. The orbitofrontal cortex receives inputs from the fusion module and information about actual and expected reinforcement from the amygdala. It compares the expected emotional learning rewards, and if they do not match, it inhibits the learning function of the orbitofrontal cortex. The inhibitory output $\mathit{O}$ in the orbitofrontal cortex is calculated as follows: $\mathit{O} = \mathit{X_{\text{am}}} \times \mathit{U}$. $\mathit{U}$ is the malleable connection weight. The connection weight $\mathit{U}$ for inhibitory output is calculated as:
	\begin{equation}
		\bigtriangleup U = \beta \left [{X_{\text{am}}} \cdot {\textstyle \sum_{0}^{j}\left (  O_{j} - {R_{\text{e}}} \right )   }   \right ], 
	\end{equation}
	where $\mathit{\bigtriangleup U}$ represents the adjustment amount of the connection weight $\mathit{U}$, $\mathit{\beta}$ is the learning rate. $\mathit{X_{\text{am}}}$ represents the comprehensive feature vector from the fusion module. $\mathit{R_{\text{e}}}$ is a reinforcement signal (Reward). $\mathit{\textstyle \sum_{j}^{}\left (  O_{j} - {R_{\text{e}}} \right )}$ is the total difference between the current output  $\mathit{O_{j}}$ and the expected reinforcement signal $\mathit{R_{\text{e}}}$. This formula indicates that for each output node, the adjustment of its connection weight $\mathit{U}$ depends on the current perceived input $\mathit{X_{\text{am}}}$and the total difference between all output nodes and the expected reinforcement signal $\mathit{R_{\text{e}}}$.
	
	\begin{equation}
		{E} =  {\textstyle \sum_{0}^{i}}{A_i} - {\textstyle \sum_{0}^{j}}{O_j},
	\end{equation}
	where $\mathit{A}$ is the output of the perception node, and $\mathit{O}$ is the output of the inhibitory node. The aforementioned formula illustrates the interaction and learning mechanism between the amygdala and orbitofrontal cortex within the BEL module.
	
	\section{Experiment}			
	\subsection{Experimental dataset}		
	This study utilized an dataset (Visual and auditory brain areas share a representational structure that supports emotion perception) publicly available on the OpenNEURO platform, which investigates the structural representations of emotional perception shared between visual and auditory brain regions, with a particular focus on audiovisual source files generated by emotional stimuli~\cite{sievers2021visual}. In this multimodal dataset, the emotional stimuli were generated using a dynamic model based on five parameters: velocity, irregularity, harmony/sharpness, the ratio of large movements, and the ratio of vertical movements. The output of the model was mapped to simple piano melodies or the motion trajectories of animated bouncing balls. During each model run, new emotional stimuli were probabilistically generated based on the current parameter settings. The animation and music components of the study each included 760 different samples, totaling 1520 samples. A total of 79 participants were involved, of which 47 were female. The output of these models is mapped through simple piano melodies or the motion trajectory of an animated bouncing ball. Participants expressed five basic emotions through manual evaluation: fear, sadness, anger, calmness, and happiness. Based on the SnowNLP sentiment analysis method, we rank the emotional categories according to their degree of positivity and calculate the final emotion score through a weighted average, followed by normalization to ensure the scores fall within the [0, 1] range, where values closer to 1 indicate more positive emotions. This research aims to explore the impact of multimodal stimuli on emotional perception and provide empirical support for the mechanisms of emotion generation.																					
	\subsection{Experimental setup}

	Our implementation details are as follows, which we aim to incorporate into the experimental procedures section. The CORnet-Z module uses a multi-convolution block architecture. It utilizes a 7$\times$7 convolution kernel with a stride of 2 and an increasing number of convolutional channels at each layer, implementing Xavier initialization to accelerate convergence. The neural simulation module models neuronal responses through dynamic equations based on voltage variations, and SpikeMonitor is used to track spike events. Model parameters include time constants $\mathit{\tau}$, input currents $\mathit{I}$, thresholds, and reset mechanisms, which influence the rate of voltage decay and spike firing. The simulation is configured by defining neuron populations and their differential equations while specifying input currents, with spike counts serving as feature inputs for subsequent models. The fusion module utilizes a multilayer perceptron architecture and the Adam optimizer (with a learning rate of 0.001) for training, which runs over 100 epochs with a batch size of 32. To prevent overfitting, L2 regularization and dropout techniques are incorporated. Experiments are conducted in a CPU environment, offering lightweight characteristics, with the potential for performance enhancement through GPU acceleration.
    \begin{figure*}[htbp]
    	\centering
    	\includegraphics[width=\linewidth]{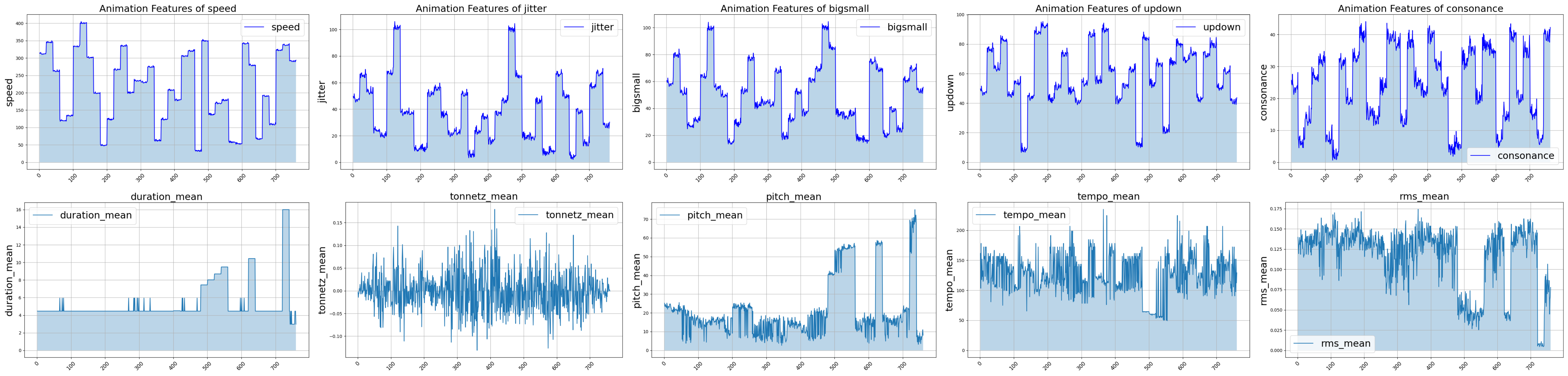}
    	\caption{In our model, auditory and visual emotional features are represented by five key acoustic parameters(Second row)—pitch, timbre, volume, tempo, and syllable length—and five key visual parameters(first row)—speed, jitter, consonance, bigsmall, and updown—extracted from animation and music.}
    	\label{fig:5}
    \end{figure*}

    As illustrated in Fig \ref{fig:5}, following the data preprocessing stage, we successfully extracted five key visual feature parameters—namely speed, jitter, consonance, bigsmall, and updown—from the animation model. These parameters are integral to the operation of the model, as they encapsulate critical aspects of the visual characteristics that influence the behavior of the animation. They will serve as essential inputs for evaluating the performance of the model, offering both quantitative and qualitative support to demonstrate its effectiveness in generating and manipulating animation features. By analyzing these parameters, we can gain deeper insights into the ability of the model to process and respond to dynamic visual stimuli, ensuring its alignment with the desired outcomes and enhancing its predictive capabilities. And after the data preprocessing stage, we also successfully extracted five essential acoustic feature parameters—pitch, timbre, volume, tempo, and syllable length—from the music model. These parameters are fundamental in capturing the intricate acoustic properties that define the musical elements and structure. They will serve as vital components for assessing the performance of the model, providing both quantitative metrics and qualitative insights into the ability of the model to analyze and generate musical features. By leveraging these parameters, we can effectively evaluate the capacity of the model to reflect the nuances of musical composition, ensuring that it produces outputs that are not only consistent with the input data but also align with the intended artistic or analytical objectives. These features will be crucial in validating the overall effectiveness of the model and its ability to process complex acoustic data accurately and meaningfully.
		
	Based on the functionality of the visual cortex module, five emotional visual features are extracted from the original emotional animation video and output as feature vectors $\mathit{X_{\text{a}}}$. Similarly, the auditory cortex module extracts five emotional auditory features from the original emotional music audio, producing feature vectors $\mathit{X_{\text{m}}}$. In the fusion module, $\mathit{X_{\text{a}}}$ and $\mathit{X_{\text{m}}}$ are combined into a comprehensive feature vector $\mathit{X_{\text{am}}}$. This vector is used as input for a recurrent neural network based on the BEL module to learn and generate emotional positivity parameter $\textit{EPP}$. The $\textit{EPP}$ is a critical metric for assessing the efficacy of our model, and we will offer a thorough explanation of it shortly.
	
	\subsection{Emotional positivity parameter}
	The $\textit{EPP}$ utilized in this study is a standardized metric designed to measure levels of emotional positivity. Inspired by the sentiment analysis tool SnowNLP, which calculates sentiment probabilities based on the emotional tendencies of words within text sentences. This parameter is then normalized to generate a value typically ranging from 0 to 1, indicating the degree of emotional inclination within the text. Our $\textit{EPP}$ adopts a similar design to the sentiment analysis tool SnowNLP~\cite{he2024novel}. Initially, it arranges manually assessed data of the experimental subjects' five emotions (fear, sadness, anger, calmness, happiness) from negative to positive. Through weighted averaging and normalization, emotional indicators are converted into values between 0 and 1, allowing for the measurement of emotional positivity in both video and audio.
	
	\begin{equation}
		\textit{EPP} = {w_{\text{f,s,a,c,h}} \cdot( E_{\text{f}} + E_{\text{s}} + E_{\text{a}} + E_{\text{c}} + E_{\text{h}})}.
	\end{equation}

    The variables $\mathit{w_{\text{f,s,a,c,h}}}$ represent the weights of fear, sadness, anger, calm, and happiness. The values of $\mathit{E_{\text{f}}}$, $\mathit{E_{\text{s}}}$, $\mathit{E_{\text{a}}}$, $\mathit{E_{\text{c}}}$, and $\mathit{E_{\text{h}}}$ are the corresponding emotion scores. Determine the $\textit{EPP}$ for each sample by applying the aforementioned formula, serving as the foundation for future experiments. Given the inherent bias in the emotion scores of the chosen dataset (e.g., if the animation displays positive emotions, the negative emotion score will be 0), we maintain consistent weights across all emotion categories when performing the weighted average. This approach helps to maximize the retention of emotional tendencies within the dataset, ensuring that the calculated emotion scores effectively reflect these tendencies without overly amplifying any specific emotion. The utilization of $\textit{EPP}$ allows for the effective measurement and comparison of emotional tendencies across various stimuli, thus offering an objective and intuitive standard for emotional assessment.
   
	\subsection{Interpretability analysis}
	\label{subsec:subsection1}
	The experiments in this study utilize pre-trained models to learn emotion features from both video and audio. For each sample, the perception node output $\mathit{A}$ and the inhibition node output $\mathit{O}$ are calculated. The model output $\mathit{E}$ is derived from $\mathit{A}$ and $\mathit{O}$, applying the inhibitory effect of the orbitofrontal cortex, and is further adjusted to generate the final $\textit{EPP}$ value. Our M-BEL and A-BEL models directly calculate the Euclidean distance percentages between the actual $\textit{EPP}$ values of all targets and the generated $\textit{EPP}$ values, then average these percentages. Subsequently, an exponential function is used to convert these averaged values into similarity percentages. For the AVF-BEL model, a weighted average of the $\textit{EPP}$ is calculated for each sample, considering the proportion of emotional features blended in both animation and audio.	The percentage Euclidean distance between the true $\textit{EPP}$ values and the generated $\textit{EPP}$ values for all targets is calculated, averaged, and then transformed into a similarity percentage using an exponential function. The subsequent analysis involves visual comparisons between the manually evaluated normalized $\textit{EPP}$ of each sample and the normalized $\textit{EPP}$ generated by the three models.
    
    \begin{figure}[h]
    	\centering
    	\includegraphics[width=\linewidth]{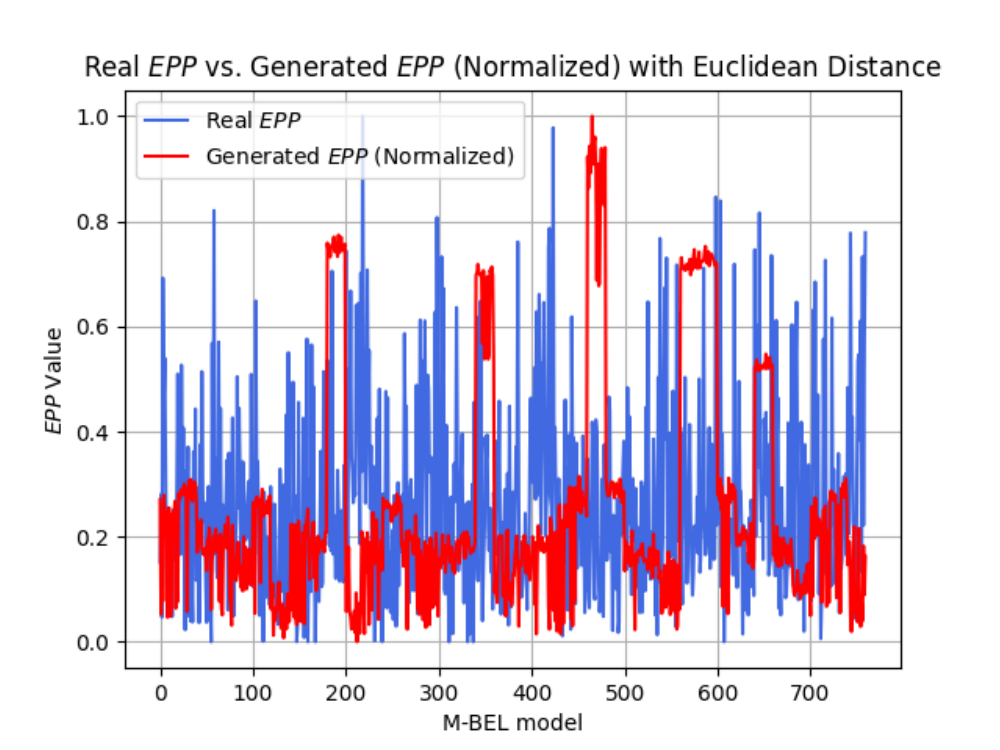}
    	\caption{The similarity comparison between the true $\textit{EPP}$ and the generated $\textit{EPP}$ in the M-BEL model.}
    	\label{fig:6}
    \end{figure}
    
	As illustrated in Fig \ref{fig:6}, it provides a visual comparison between the manually evaluated normalized $\textit{EPP}$ of each sample and the normalized $\textit{EPP}$ generated by the M-BEL model. Due to the limitations of the music data features in the dataset and the incompleteness of the M-BEL model, there is a significant difference between the $\textit{EPP}$ generated by the M-BEL model and the actual $\textit{EPP}$ obtained through manual evaluation.
	
	\begin{figure}[h]
		\centering
		\includegraphics[width=\linewidth]{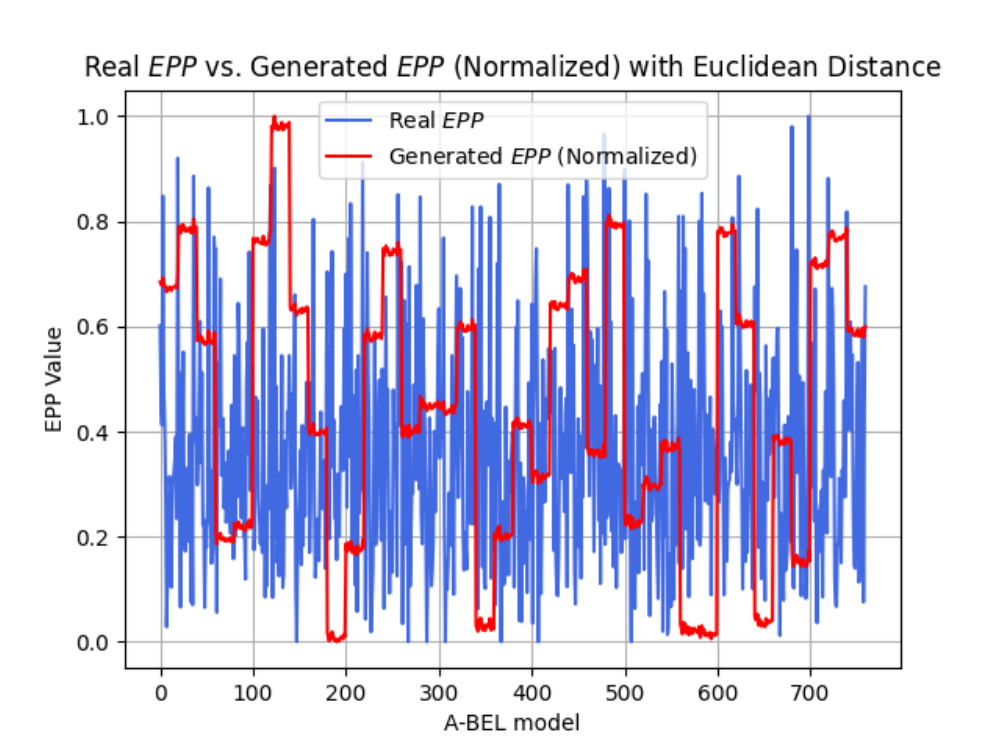}
		\caption{The similarity comparison between the true $\textit{EPP}$ and the generated $\textit{EPP}$ in the A-BEL model.}
		\label{fig:7}
	\end{figure}

	As illustrated in Fig \ref{fig:7}, it provides a visual comparison between the manually evaluated normalized $\textit{EPP}$ of each sample and the normalized $\textit{EPP}$ generated by the A-BEL model. Due to the limitations of the animation data features in the dataset and the incompleteness of the A-BEL model, there is also a considerable difference between the $\textit{EPP}$ generated by the A-BEL model and the actual $\textit{EPP}$ obtained through manual evaluation. However, compared to the M-BEL model, it demonstrates a significant improvement in performance. This is also consistent with the observed finding that visual features contain more emotional information than auditory features, with visual features being more likely to trigger emotional changes and other physiological phenomena.
    	
    \begin{figure}[b]
    	\centering
    	\includegraphics[width=\linewidth]{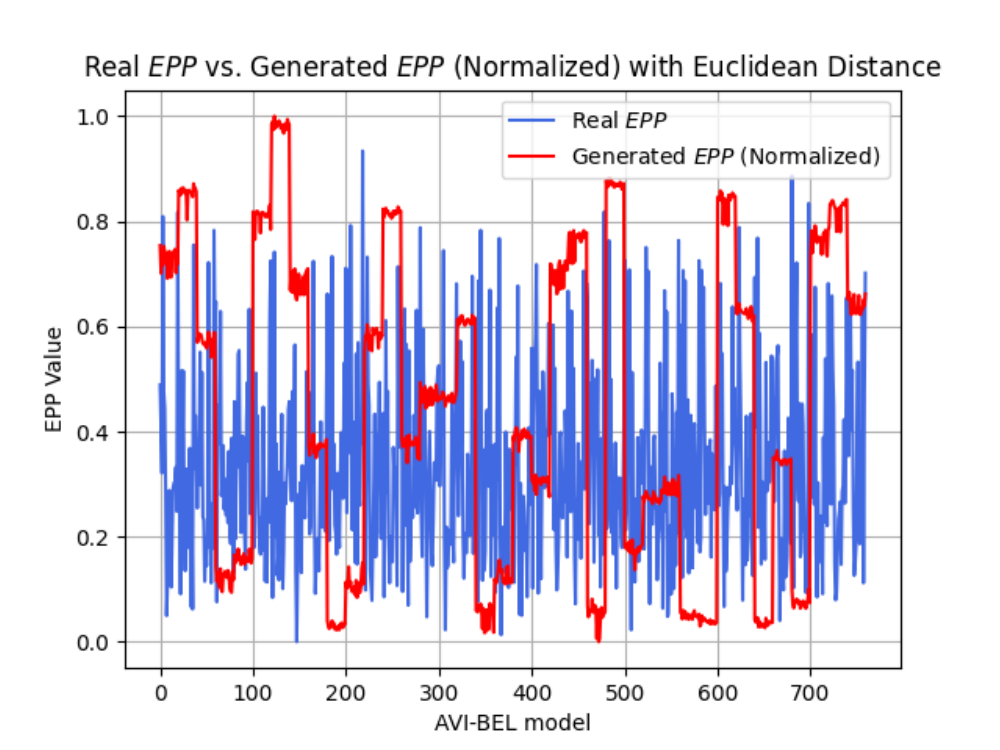}
    	\caption{The similarity comparison between the true $\textit{EPP}$ and the generated $\textit{EPP}$ in the AVF-BEL model.}
    	\label{fig:8}
    \end{figure}
    
	As illustrated in Fig \ref{fig:8}, it provides a visual comparison between the manually evaluated normalized $\textit{EPP}$ of each sample and the normalized $\textit{EPP}$ generated by the AVF-BEL model. Although limited by the constraints of the data features in the dataset, the AVF-BEL model is more refined than the M-BEL and A-BEL models. The difference between the $\textit{EPP}$ generated by the AVF-BEL model and the actual $\textit{EPP}$ obtained through manual evaluation has been further reduced. This is also consistent with the physiological responses of the human brain, where emotional changes triggered by the simultaneous interaction of visual and auditory features are stronger than those induced by a single modality.
	 
   As illustrated in Fig \ref{fig:1} and Fig \ref{fig:2}, by integrating the content from Sections \ref{subsec:3.1} and \ref{subsec:subsection0}, we conducted an interpretability analysis that clearly reveals the roles and interrelationships of the four neural mechanisms and framework modules in the emotional learning and generation process. This analysis provides a direct observation of how the structure and function of the model map onto different regions of the emotional pathways of the brain and their underlying biological mechanisms. Furthermore, through the progressive presentation of three comparative experiments, we validate the significant advantages of the audiovisual fusion emotion learning and generation model in emotional expression. The results of these comparisons demonstrate that the multimodal learning mechanism, which integrates visual and auditory information, significantly enhances the accuracy and diversity of emotional generation, aligning with the combined effects of visual and auditory stimuli. This finding underscores the effectiveness and reliability of the model in simulating the multimodal emotional processing capabilities of the brain. Through the analysis of these experimental results, we further confirm the interpretability of the model, establish the core role of audiovisual fusion in emotion learning and generation, and provide a theoretical foundation for the optimization and development of future multimodal emotion generation models.
	
	\subsection{Lightweight analysis}
	Our AVF-BEL model uses a multi-convolutional block architecture that combines neural simulation and fusion modules. This structure allows the model to accurately simulate neuronal responses with dynamic equations and use convolutional layers for spatial feature extraction. Although the convolutional layers contain a large number of parameters, the model complexity is managed by using 7$\times$7 convolutional kernels and gradually increasing the number of convolutional channels. Experiments conducted in a CPU environment emphasize lightweight characteristics, allowing the model to operate effectively in resource-constrained settings. In contrast, traditional deep learning models typically utilize fixed convolutional network architectures that often entail high parameter counts and computational demands, usually requiring GPU acceleration to meet real-time processing needs, especially when handling large-scale video and audio data, where computational resource consumption becomes significantly pronounced.
		
		\begin{table*}[htbp] 
		\centering
		\caption{Comparative analysis of the AVF-BEL model with deep learning multimodal models: a case study of Multimodal Transformers} 
		\begin{tabular}{lcc}
			\toprule 
			Comparison Items & AVF-BEL & Multimodal Transformers \\
			\midrule 
			Model Depth & 2 layers & 12-24 layers, uses self-attention \\
			Parameters & Small(Tens of)& Large(Millions to billions) \\ 
			Computational Complexity & Linear level(O(n)) & Square level((O(n²))) \\ 
			Memory Footprint & CUP level(Less than 8GB) & needs GPU/TPU (Above 16GB)\\ 
			Dataset Size & Small(Below 2 GB) & Large(More than 10 GB) \\ 
			Self-Attention & Not used, relies on lightweight models & Core mechanism \\
			Application & Low-resource, fast response & High-precision, complex tasks \\ 
			\bottomrule 
		\end{tabular} 
		\label{tab:1} 
	\end{table*}
	
    The table \ref{tab:1} provides a comprehensive comparison between the AVF-BEL model and multimodal transformers, highlighting significant differences across several key features\cite{yoon2023multimedia}. The AVF-BEL model has a shallower architecture, consisting of only two layers, making it suitable for rapid responses in low-resource environments. Its small parameter count (only a few dozen) results in linear computational complexity, which enhances its efficiency and memory usage, allowing it to run effectively on central processing units (CPUs) and handle datasets smaller than 2 GB. In contrast, multimodal transformers feature a deeper architecture, typically consisting of 12 to 24 layers, and use self-attention mechanisms to effectively capture complex relationships in the data. They are designed to process large datasets exceeding 10 GB. The parameter count for these models is substantial (often in the millions to billions), and their computational complexity is quadratic, necessitating the use of graphics processing units (GPUs) or tensor processing units (TPUs). Consequently, multimodal transformers excel in high-precision and complex tasks. In summary, the AVF-BEL model is more suitable for resource-constrained applications, while multimodal transformers demonstrate superior expressive capabilities and efficiency in handling large-scale data, catering to diverse application requirements.
    
    \begin{figure}[b]
    	\centering
    	\includegraphics[width=\linewidth]{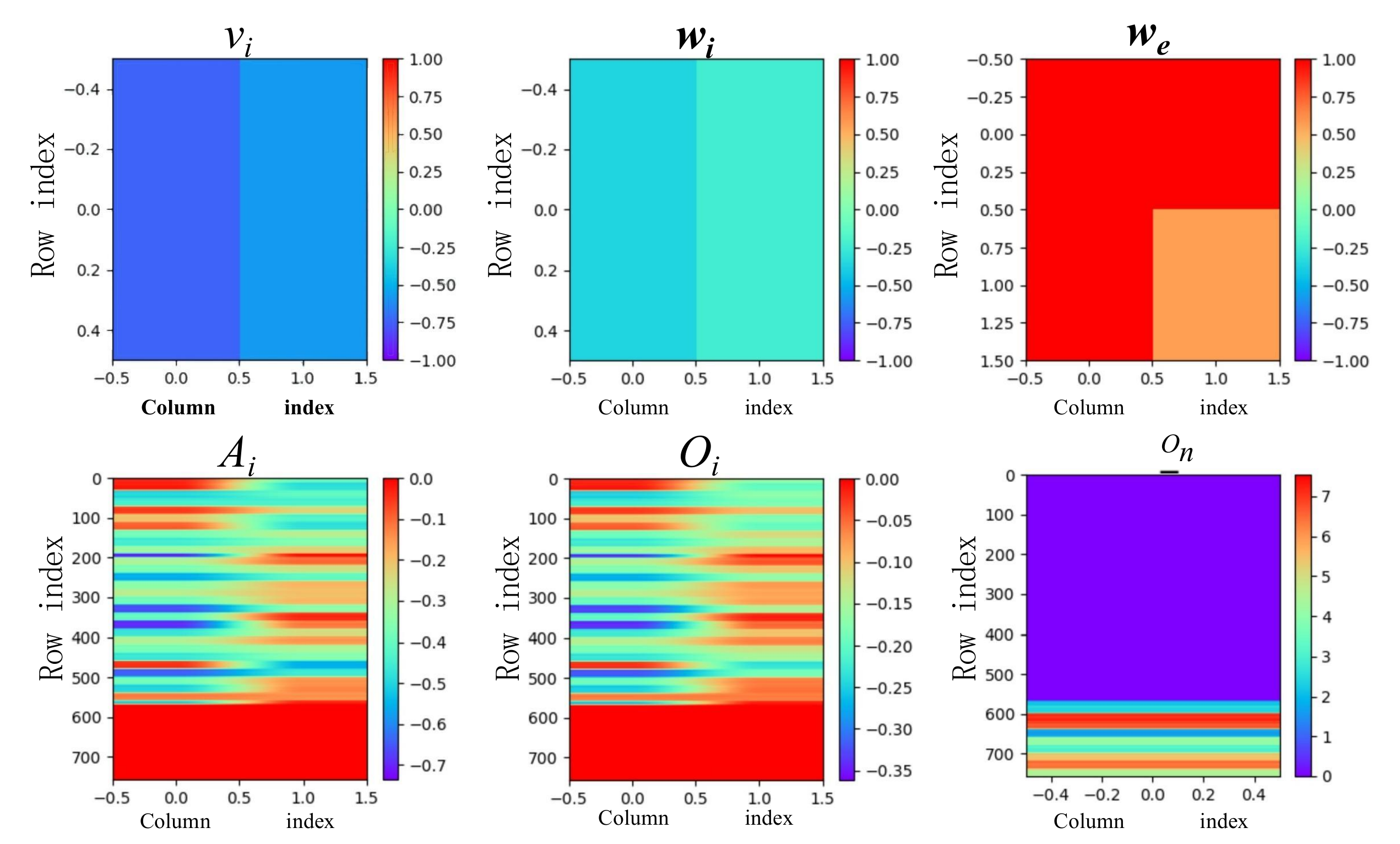}
    	\caption{Heatmap of model weight matrices: visualization of input feature processing, excitatory and inhibitory activation, and prefrontal cortex regulation in neural simulation.}
    	\label{fig:9}
    \end{figure}
    
	The weight matrices $\mathit{v_{\text{i}}}$,$\mathit{w_{\text{i}}}$, and $\mathit{w_{\text{e}}}$ hold pivotal roles within both the neural network and the simulation model. They are utilized respectively for processing input features, computing activation values of excitatory and inhibitory systems, and simulating the regulatory function of the prefrontal cortex. The weight values and heatmaps reflect not only the significance of input features within the model but also the response patterns of various systems. $\mathit{A}$ represents the output of the excitatory node, calculated based on the input features and weights $\mathit{v_{\text{i}}}$. During training, it is used to evaluate the contribution of excitatory inputs to the model. $\mathit{O}$ represents the output of the inhibitory node, calculated based on the input features and weights $\mathit{w_{\text{i}}}$. During training, it is used to assess the impact of inhibitory outputs on the model. The variable $\mathit{o_{\text{n}}}$ normalizes the predicted output of the model, scaling its values to the range of 0 to 1 using the minimum and maximum values of the target variable. This normalization makes the resulting values easier to compare and visualize, as they fall within a standardized range (typically 0 to 1). This standardization makes comparisons between different features more intuitive and meaningful.

    The heatmap of specific parameters in the model, as depicted in Fig \ref{fig:9}, reveals that many of the weight values are approaching zero, indicating a high degree of sparsity within the parameter set. This sparsity is a significant characteristic of the model, as it leads to a reduced number of active parameters, which in turn contributes to a more efficient representation. The presence of sparsity helps the model mitigate the risk of overfitting and avoids the entrapment in local minima during the training process, which is often a challenge in complex models with large parameter spaces. By promoting the activation of only the most relevant parameters, the model enhances its ability to generalize from the training data to unseen data, ensuring more robust and scalable performance. These observations, derived from the weight parameter heatmaps, strongly suggest that the AVF-BEL model possesses lightweight attributes, with a streamlined parameter structure that supports both computational efficiency and improved generalization.
    
    In summary, the AVF-BEL model demonstrates significant lightweight advantages, as evidenced by its shallow architecture, sparse parameter distribution, and low computational complexity. By using a multi-convolutional block design with 7×7 convolutional kernels and progressively increasing convolutional channels, the model maintains high efficiency while minimizing the number of active parameters. This sparsity not only reduces memory and computational demands, enabling effective operation in resource-constrained environments, but also enhances generalization by preventing overfitting and avoiding local minima during training. As such, the AVF-BEL model stands out for its balance between performance and resource efficiency, making it well-suited for low-resource applications without compromising its functional capabilities.
   
	\subsection{Ablation study}
	\label{subsec:subsection2}
	
	In addition to using similarity metrics to assess the generated Emotion Parameter Profile $\textit{EPP}$ values, we enhance the evaluation by converting both the normalized generated $\textit{EPP}$ values and true $\textit{EPP}$ values into binary labels (0 or 1). This transformation frames the task as a classification problem, enabling the use of classification metrics such as precision, recall, and F1-score. These metrics effectively evaluate the performance of the model in distinguishing between generated and true $\textit{EPP}$. Precision measures the proportion of true positives among predicted positives, recall assesses the proportion of true positives among actual positives, and the F1-score balances both metrics. This approach provides a more comprehensive validation of the effectiveness and accuracy of the model, offering a clearer assessment of its ability to generate $\textit{EPP}$ values that align with true emotional parameters.
		
	\begin{table*}[htbp]
		\centering
		\caption{Comparative assessment of experimental modules in BEL, M-BEL, A-BEL, and AVF-BEL models}
		\begin{tabular}{lccccc}
			\toprule
			Module & BEL(m)& BEL(a) & M-BEL & A-BEL & AVF-BEL\\
			\midrule
			Visual Cortex & - & - & - & $\checkmark$ & $\checkmark$ \\
			Auditory Cortex & - & - & $\checkmark$ & - & $\checkmark$ \\
			Fusion module & - & - & - & - & $\checkmark$ \\
			Emotional Learning & $\checkmark$ & $\checkmark$ & $\checkmark$ & $\checkmark$ &$\checkmark$\\
			\midrule
			Precision & 0.57& 0.55 & 0.58 & 0.65 & 0.79 \\
			Recall & 0.44& 0.68 & 0.47 & 0.69 & 0.77 \\
			F1-score & 0.50& 0.61 & 0.52 & 0.67 & 0.78 \\
			Average similarity & 48.74\% & 65.08\% & 49.26\% & 67.48\% & 77.69\% \\
			\bottomrule
		\end{tabular}
		\label{tab:2}
	\end{table*}
	
	As shown in Table \ref{tab:2}, The BEL are implementations of the traditional brain emotion model based on recurrent neural network technology. The traditional BEL model has not adequately considered the information processing of the sensory cortex; therefore, it is necessary to input audio and video features separately to evaluate the performance of the model on the dataset. As shown in the table, the BEL model achieved an accuracy of 0.57, recall of 0.44, and an F1 score of 0.50 on the music data in the dataset. Additionally, the average similarity between the $\textit{EPP}$ generated by the BEL model and the actual $\textit{EPP}$ obtained through manual evaluation is 48.74\%.  The BEL model achieved an accuracy of 0.55, recall of 0.68, and an F1 score of 0.61 on the animation data in the dataset. Additionally, the average similarity between the $\textit{EPP}$ generated by the BEL model and the actual $\textit{EPP}$ obtained through manual evaluation is 65.08\%. The M-BEL is an enhanced BEL model that reinforces auditory cortex processing, consisting of an auditory cortex module and an emotional learning module.The M-BEL model attained a precision of 0.58, a recall of 0.47, and an F1-score of 0.52. The average similarity, calculated as the average Euclidean distance transformed into a percentage, between the $\textit{EPP}$ generated by the M-BEL model and the actual $\textit{EPP}$ obtained through manual assessment is 49.06\%. The A-BEL is an improved BEL model that enhances visual cortex processing, comprising a visual cortex module and an emotional learning module. The A-BEL model, which is a visual emotion model consisting of visual cortex modules and emotion learning modules, achieved a precision of 0.65, a recall of 0.69, and an F1-score of 0.67. The average similarity, calculated as the average Euclidean distance transformed into a percentage, between the $\textit{EPP}$ generated by the A-BEL model and the actual $\textit{EPP}$ obtained through manual assessment is 65.08\%. By comparing these data, we found that the M-BEL and A-BEL models, which incorporate the auditory cortex and visual cortex modules into the BEL model, show some improvement in the emotional generation similarity for both music and animation data.
	
	Our ultimate AVF-BEL model integrates visual cortex modules, auditory cortex modules, fusion modules, and emotion learning modules. The AVF-BEL model, which implements the BEL module using a RNN, achieved an accuracy of 0.79, a recall of 0.77, and an F1 score of 0.78. The average similarity, calculated as the average Euclidean distance transformed into a percentage, between the $\textit{EPP}$ generated by the AVF-BEL model and the actual $\textit{EPP}$ obtained through manual assessment, is 77.69\%. The visual cortex module offers more comprehensive information compared to the auditory cortex module, thereby contributing to the superior performance of the A-BEL model over the M-BEL model. The fusion module integrates visual and auditory cortical inputs to generate audiovisual fused stimuli, further enhancing the performance advantage of the AVF-BEL model. The replacement and addition of bio-inspired modules demonstrate a clear enhancement in model performance. These experimental results validate the significant improvement in emotion generation when using fused visual and auditory stimuli. Furthermore, the study highlights the superior performance of the AVF-BEL model in computational metrics, particularly its advantage in neuroanatomical alignment, offering novel insights and methodologies for research in emotion generation.

	The ablation study results demonstrate the significant impact of each module within the AVF-BEL model. The addition of the visual cortex and auditory cortex modules in the M-BEL and A-BEL models, respectively, enhances the processing of sensory inputs, leading to improvements in precision, recall, and F1-score. The AVF-BEL model, which integrates both visual and auditory cortex modules along with a fusion module, outperforms all other models across all metrics. The fusion module, which integrates visual and auditory inputs, is key to enhancing the ability of the model to generate more accurate emotional responses, achieving an average similarity of 77.69\%, well beyond the performance of the individual M-BEL and A-BEL models. These results emphasize the importance of integrating multisensory stimuli to enhance emotion generation and further validate the advantages of the AVF-BEL model in neuroanatomical alignment and interpretability, providing valuable insights for future research in brain-inspired emotional computation.
	
	\section{Conclusion}
	\label{sec:conclusion}
	The AVF-BEL model presented in this study is designed to replicate the neural pathways associated with emotion learning and generation, specifically focusing on how visual and auditory perceptions are integrated and transformed into emotional stimuli in the brain. By simulating the biological neural mechanisms involved in emotional processing, this model seeks to advance the development of lightweight, neuroanatomically aligned methods for emotion learning and generation. The model not only mimics the cortical processing of emotional features related to visual and auditory stimuli but also introduces an additional level of sophistication by simulating the integration of these sensory modalities in the anterior superior temporal gyrus, an area of the brain known for its role in multisensory integration. This biomimetic approach allows the AVF-BEL model to more accurately reflect how the brain processes and combines visual and auditory emotional information, ultimately enabling the generation of emotion parameters that are more robust and nuanced. Furthermore, by incorporating these biomimetic functionalities, the AVF-BEL model achieves enhanced interpretability, providing clearer insights into the mechanisms underlying emotion generation. The lightweight design of the model ensures computational efficiency, making it suitable for applications in resource-constrained environments. In future research, we aim to further improve the neuroanatomical alignment of the AVF-BEL model and explore the integration of additional biomimetic modules, such as those representing the hypothalamus and the cingulate gyrus. These advancements will likely enhance the ability of the model to simulate emotional regulation and processing at a more comprehensive level, potentially extending its applicability to a broader range of real-world scenarios and improving its alignment with more complex brain-based emotion generation mechanisms.

	\newenvironment{acks}{\section*{Acknowledgements}}{}
\begin{acks}
	The authors gratefully acknowledge the financial support provided by the Natural Science Foundation of Hunan Province (No.2024JJ6190), the Open Project of Xiangjiang Laboratory (No.23XJ03009), Xiangjiang Laboratory major program subproject (No.22XJ01001-2), the "Digital Intelligence +" interdisciplinary research project of Hunan University Of Technology and Business (No.2023SZJ19), Scientific research project of Hunan Provincial Department of Education (No.22B0646).
\end{acks}

\bibliographystyle{elsarticle-num}
\bibliography{reference-base}

\balance 

\end{document}